# An efficient HTS electromagnetic model combining thin-strip, homogeneous and multi-scale methods by *T-A* formulation

Lei Wang and Yan Chen

State Key Laboratory of Surface Physics and Department of Physics, Fudan University, Shanghai, 200433, China

E-mail: wangleifudan@fudan.edu.cn and yanchen99@fudan.edu.cn

**Abstract**

This study presents an HTS electromagnetic model combining the thin-strip, homogeneous and multi-scale methods using *T-A* formulation. In particular, we build the thin strips as both the analyzed HTS tapes and the boundaries of the homogeneous bulks where the non-analyzed tapes are merged. Thus, the coil geometry is re-constructed with several bulks, but the bulks' boundaries and domains are tackled with different electromagnetic properties, and solved by *T* and *A* formulations, respectively. Firstly, we introduce the modeling process and highlight the differences and advantages over the previous models. Then, the accuracy of the proposed model is validated by comparing the results with those from the reference model based on a 2000-turn coil. The distributions of normalized current density, magnetic flux density and hysteresis losses from the two models are highly consistent, and the error of the total loss is less than 1%. Besides, the proposed model is the most time-saving among all the advanced models. Furthermore, the model can be applied in 3D simulations, and the high accuracy and efficiency are validated by simulating a 50-turn racetrack coil. The proposed method provides a feasible approach to simulating coils with many stacked tapes, and we will continue exploring more applications in solving HTS systems with complex geometries.

## I. INTRODUCTION

Finite element (FE) numerical model has been a powerful tool for understanding the electromagnetic behaviors of high temperature superconductors (HTS). Based on the state variables to be solved, different FE models are defined: *A-V* [1,2], T-Ω [3,4] and *H* [5,6] formulations. They have been widely employed to simulate HTS coils and estimate AC loss, mainly due to their easy realization with commercially available software packages. Especially the *H* formulation became quickly popular and is now the de facto standard in the applied community [7].

In the electromagnetic modeling of rare-earth-barium-copper-oxide (ReBCO) tapes, the highly nonlinear *E-J* power relationship and the high aspect ratio impede the traditional numerical models to be applied for simulating large-scale coils. The former factor makes the calculation challenging to converge, and the latter leads to large numbers of mesh nodes and degrees of freedom to be solved. To address the issue, two main approaches are proposed according to the factors causing difficulties: eliminating the high aspect ratio or reducing the simulations of ReBCO tapes with *E-J* power law.

The high aspect ratio of superconducting tapes can be changed by expanding or eliminating their thicknesses. Accordingly, the approximations of anisotropic bulks and thin strips are developed. Based on the anisotropic homogeneous-medium approximation, Zermeno *et al*. employed the homogenization method to simulate a stack of ReBCO tapes with anisotropic bulks using *H* formulation [8]. Later, the *H* homogeneous method was extended to the three-dimensional (3D) simulation of a racetrack coil [9] and a two-dimensional (2D) axisymmetric model of solenoid coils [10]. By using the thin strip approximation, the thicknesses of superconducting layers are supposed to be zero. With the reduced-dimensionality tapes built, *T-A* formulation was proposed, in which the current vector potential *T* and the magnetic vector potential *A* are used to calculate the tape's current density and the spatial magnetic field, respectively. The *T-A* thin-strip method was employed to simulate 3D HTS models [11,12] and 2D large-scale stacks [13]. Later, the *T-A* formulation was applied in the bulk domains for 2D and 3D simulations [14–16]. The degrees of freedom and computing time are reduced significantly with bulks or thin strips built instead of the actual-geometry tapes. Besides, the *T-A* homogeneous model is more efficient than the one solved by the *H* formulation [17]. Because the *T* and *A* potentials are effectively coupled, which enable a straightforward calculation of current density and magnetic field, respectively. Despite the deficiency that the magnetic field parallel to the tape's surface cannot be tackled correctly because of the changed thicknesses, the models provide reliable simulations for the cases where the influence of the parallel magnetic field can be negligible [10,14].

The idea of reducing simulations of HTS tapes can be found in the multi-scale method proposed by Queval *et al* [18]. The model solves the magnetic field of the full coil built with tapes' actual geometries by *A* formulation first (called *coil submodel*), and then calculates hysteresis losses of several HTS tapes at significant positions (called *analyzed tapes*) by *H* formulation with the obtained magnetic field (called *single-tape submodel*). The losses in the non-analyzed tapes are estimated by interpolation. Infinite-array approximation [18,19] and iterative method [14,20] are employed to enhance the magnetic field estimation, which determines the accuracy of the hysteresis loss calculation. Since the calculation of HTS system is broken up to solving single tapes, the method can be applied in simulating giant coils. However, the two sub-models are implemented in two individual computing files, and additional procedures are needed to control the calculations. A new approach called the densification method also addresses the analysis of stacks with a reduced number of tapes (called *densified tapes*) [17]. A densified tape merges a given number of tapes in the densification process, and the merged tapes are erased. Accordingly, the transport current carried in the densified tapes is the sum of currents in the merged tapes. Nevertheless, the densified tapes preserve the single-tape's original geometry (referring to Figure 8 in [17]).

Recently, combination models have been studied, in which two or more advanced methods (homogenization, thin strip, multiple scales, and densification) are employed in one numerical model to explore more feasible approaches on simulating large-scale HTS systems efficiently. Berrospe-Juarez *et al* combined the thin-strip and multi-scale methods by *T-A* formulation [14], where a significant enhancement was made on the multi-scale method, that is, the calculations of the analyzed tapes' current densities and the coil's magnetic field are implemented simultaneously in one computing file. And the model is referred to *simultaneous multi-scale model* [14]. We ever introduced the homogenization technique into the multi-scale method to simplify the coil sub-model building. Based on the simplification, we proposed the multi-dimension method capable of simulating 3D HTS coil models for the first time [21]. In the study [17], Berrospe-Juarez *et al* summarized 14 models using *H* or *T-A* formulations, in which the combination of two or more methods presented various new models. In principle, the more advanced methods that are employed, the faster the simulation will be. However, as we can see in the models comparison (referring to Table 3 in [17]), the *T-A* homogeneous model is the most efficient one, in which only the homogenization method is employed. Furthermore, the computing time of the *T-A* model that combines thin strips, homogeneous bulks, and multi-scale method is much longer than that of the homogeneous model in which bulk domains are solved by *T-A* formulation directly. In the combination model, the thin strips (including the analyzed and non-analyzed tapes) and homogenized bulks are constructed separately, making the calculation complicated.

In this paper, we propose a new method that combines the thin-strip, homogeneous and multi-scale methods in one numerical model by using *T-A* formulation. Section II introduces the modeling strategy, and the differences and advantages over the previous models. In Section III, a coil of 2000-turn stacked tapes is simulated with different models to validate the accuracy and efficiency of the proposed one. The application in the 3D simulation is presented based on a racetrack coil in Section IV. Conclusions are given in Section V.

## II. METHODOLOGY: *T-A* SIMULTANEOUS MULTI-SCALE HOMOGENEOUS MODEL

### A. Modeling strategy

**Figure 1** presents the modeling strategy of the proposed model. Firstly, we select several analyzed tapes at significant positions based on the idea of multi-scale method. And then, the analyzed tapes are built with thin strip approximation. With those strips as boundaries, the coil geometry can be divided into several bulks. Consequently, the analyzed tapes exist as the boundaries of the bulks, and the non-analyzed tapes are totally merged into the bulk domains. Equivalently, the building process can be considered as that the homogenization technique is employed to build the whole coil geometry with bulks, and the bulks' edges parallel to the tape's surface are treated as HTS tapes with the thin strip approximation.

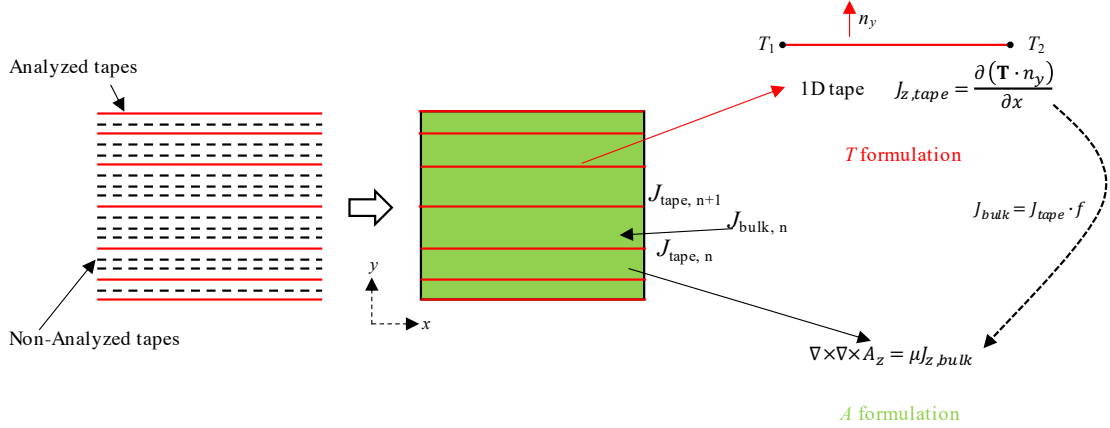

**Figure 1**. Modeling strategy of the proposed model. $J_{\text{tape, n}}$ and $J_{\text{tape, n+1}}$ represent the current density in the two adjacent analyzed tapes. $J_{\text{bulk, n}}$ represents the current density in the bulk domain between the tapes n and n+1. $f$ stands for the ratio of the superconducting layer's thickness to the tape's.

As shown in **Figure 1**, the current density in the analyzed tapes (or the bulks' parallel edges) is calculated by the current vector potential **T**: $\mathbf{J} = \nabla \times \mathbf{T}$. Since **J** is constrained to flow within the strip and **T** is normal to the conductor's wide face at each point, **T** can be written as $T \cdot \vec{n}$, where $\vec{n} = \begin{bmatrix} n_x \\ n_y \\ n_z \end{bmatrix}$ is the locally normal vector of the strip, and $T$ is the magnitude of the current vector potential. In this case, 2D planar geometry is used, and the tapes are 1D lines, currents flow in the

longitudinal direction only ($J_z$). The governing equations are

$$J_{z,tape} = \frac{\partial(\mathbf{T} \cdot n_y)}{\partial x} \tag{1}$$

$$-\frac{\partial}{\partial x}\rho\frac{\partial(\mathbf{T} \cdot n_y)}{\partial x} = -\frac{\partial B_y}{\partial t} \tag{2}$$

where $J_{z,tape}$, $\rho$, and $B_y$ represent the current density, resistivity, and magnetic flux density in the analyzed tape, respectively. The applied transport current $I_a$ in each tape is imposed by setting the boundary conditions for $\mathbf{T}$ ($T_1$ and $T_2$) shown in **Figure 1**. And the values of $T_1$ and $T_2$ must satisfy the relationship

$$I_a = (T_1 - T_2)\delta \tag{3}$$

where $\delta$ is the real thickness of the superconducting layer. Usually, $T_1$ or $T_2$ is set to zero, and the other one is $I_a/\delta$.

With the current densities in the analyzed tapes, the current densities distributed in the bulk domains can be characterized by interpolation. Linear interpolation is used to obtain the current density in the bulk domain *n*:

$$J_{bulk,n} = (F_1(y) \cdot J_{tape,n} + F_2(y) \cdot J_{tape,n+1}) \cdot f \tag{4}$$
$$F_1(y) = 1 - (y - y_n)/(y_{n+1} - y_n) \tag{5}$$
$$F_2(y) = (y - y_n)/(y_{n+1} - y_n) \tag{6}$$

where $J_{tape,n}$, $J_{tape,n+1}$, and $J_{bulk,n}$ represent the current density in the tape n, tape n+1, and the domain between them, respectively. $F_1$ and $F_2$ are linear functions varying with *y* between $y_n$ and $y_{n+1}$. $y_n$ and $y_{n+1}$ represent the coordinates of the edges n and n+1, respectively. *f* is the ratio of the HTS layer's thickness to the practical tape's thickness. With the current densities in all the bulk domains, the magnetic field of the whole coil can be estimated by solving *A* formulation:

$$\nabla \times (\nabla \times \mathbf{A}) = \mu \mathbf{J}_{bulk} \tag{7}$$
$$\mathbf{B} = \nabla \times \mathbf{A} \tag{8}$$

where $\mathbf{A}$ is the magnetic vector potential, $\mu$ is the magnetic permeability, and $\mathbf{B}$ is the magnetic flux density. Thus, the $B_y$ in equation (2) is obtained. In addition, the *E-J* power law is employed to simulate superconducting tapes, and the resistivity $\rho$ in equation (2) can be expressed as

$$\rho = \frac{E_c}{J_c}\left|\frac{J_z}{J_c}\right|^{n-1} \tag{9}$$

where $E_c$ is the threshold electric field used for defining critical current in measurement, $J_c$ is the critical current density, and *n* value defines the steepness of the transition from superconducting to normal state [22]. The electric field $E_z = \rho \cdot J_z$. The hysteresis loss in one tape can be calculated:

$$Q[\text{W/m}] = \frac{2}{T}\int_{T/2}^{T}\int_L (E_z \cdot J_z \cdot \delta)dldt \tag{10}$$

where L and $\delta$ are the width and thickness of a tape, respectively. With the losses of the analyzed tapes calculated, the losses in the non-analyzed tapes can be estimated by interpolation.

The *T-A* formulation is implemented with the Magnetic Field modules coupled with PDE module in lower dimensions in COMSOL. General Extrusion operators built-in COMSOL are used to call the $J_{tape}$ in equation (4) to realize the simultaneous multi-scale method.

**B. Differences and advantages over the previous models**

In the proposed model, thin strips are solved by *T* formulation as the conventional *T-A* thin-strip model, but those strips are no longer imposed with surface current density in the *A* formulation to calculate the magnetic field. Since the current densities in the bulk domains cover the carrying currents of the whole coil. Due to the characterization that all the magnetic fields are produced by the bulk domains, no magnetic field distortions appear. The distortions occurred in the *T-A* simultaneous multi-scale homogeneous model built in [17], where the tapes and bulks are built separately. In order to move the distortions in the magnetic field produced by the homogeneous tapes away from the analyzed tapes, the non-analyzed tapes adjacent to the analyzed tapes kept their original geometries and were not merged in bulks [17].

Both the *T-A* homogeneous model and the proposed one build the coils with bulks only. The difference is that the *T* formulation is applied in the bulk domains in the former, while thin strips are still solved by *T* formulation in the latter. Advantages of the thin-strip modeling include that the normal vector $\vec{n}$ can be easily determined, and no extra Neumann boundary conditions are needed. While these issues have to be considered in the *T-A* homogeneous model [14,15]. Besides, based on the idea of multi-scale method, the proposed model can build the coil with fewer bulks if the analyzed tapes and the interpolation method are appropriately chosen.

The proposed model presents a combination of thin strips, homogenization technique and multi-scale method, but the thin strips exist as both the analyzed tapes and the boundaries of the bulks. Thus, we still refer to it as the *T-A* simultaneous multi-scale homogeneous (*T-A* SMSH) model.

## III. CASE STUDY FOR VALIDATION

### A. Case study

A coil made of 2000 turns is simulated, which has been previously studied [17,18,23]. The coil geometry with individual tapes is shown in **Figure 2**, and detailed parameters are listed in **Table 1**. All the numerical models consider the superconducting layers only within the tapes and simulate 1/4 part of the coil for simplification based on the symmetry.

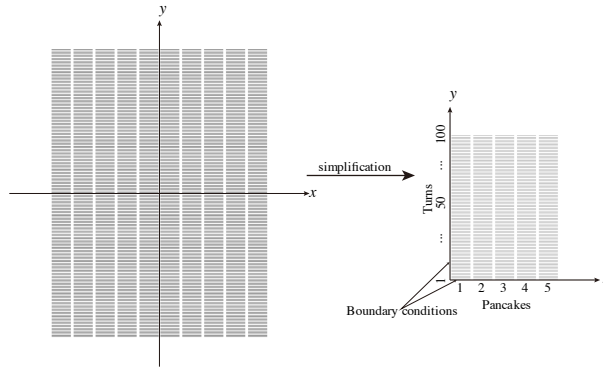

**Figure 2**. The target coil simulated in the study.

**Table 1**. Geometric parameters of the coil and tape.

| Parameter | Value |
| --- | --- |
| Number of pancakes | 10 |
| Turns per pancake | 200 |
| HTS layer width | 4 mm |
| HTS layer thickness | 1 um |
| Winding tape width | 4.4 mm |
| Winding tape thickness | 293 μm |

The Kim-like model [24] is used to describe the anisotropic dependence of the HTS tapes' critical current density on the magnetic field, and $J_c$ is defined by

$$J_c(B) = \frac{J_{c0}}{\left(1 + \frac{\sqrt{k^2 B_\parallel^2 + B_\perp^2}}{B_0}\right)^\alpha} \quad (11)$$

where $J_{c0}$, $B_0$, $k$ and $\alpha$ are material parameters. $B_\parallel$ and $B_\perp$ represent the magnetic flux density components parallel and perpendicular to the wide surface of the tape, respectively. The parameters of equations (9) and (11) are listed in **Table 2**.

Table 2. Electromagnetic parameters of the HTS tapes.

| Parameter | Value |
| --- | --- |
| $E_c$ | 1E-4 V/m |
| $n$ | 38 |
| $J_{c0}$ | 2.8E10 A/m² |
| $k$ | 0.29515 |
| $\alpha$ | 0.7 |
| $B_0$ | 42.65 mT |

The calculation accuracy of the proposed *T-A* SMSH model is validated by comparing the electromagnetic results with those from the *H* reference model. Besides, the *H* and *T-A* full models, the *H* and *T-A* homogeneous models are solved to compare the computing time. Both the *H* reference and *H* full models simulate the coil built with all the tapes in their actual 2D geometries, but the former is solved with finer meshes to keep the calculation accurate as a reference, in which the 4-mm width of a tape is meshed with 100 elements. The *H* full model and the others are meshed with 60 elements along the 4-mm width. The *H* and *T-A* homogeneous models build each stack with 10 bulks distributed logarithmically [18,23]. While five tapes with the number of {1 70 91 98 100} are selected and built as 1D lines in each stack in the *T-A* SMSH modeling. Geometries of the *H* and *T-A* homogeneous and the *T-A* SMSH models are shown in **Figure 3**. The thickness of each bulk is meshed with 1 element in both geometries. Although the bulks in the *T-A* SMSH model can be meshed with multiple elements along their thicknesses, we found that the solution is almost no difference with one meshing element.

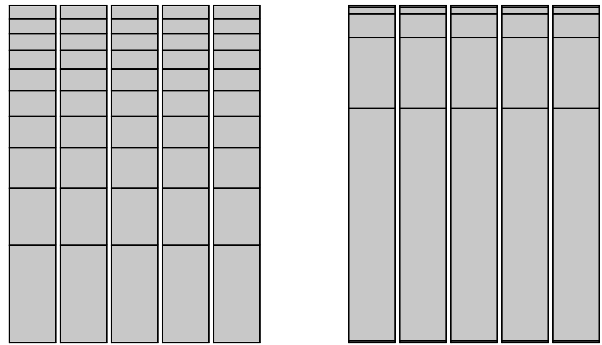

**Figure 3**. The coil geometries built in the *H* and *T-A* homogeneous models (left) and the *T-A* SMSH model (right).

In the *H* models, the linear element function is used for solving, while in the *T-A* models, the linear and quadratic element is used for solving the *T* and *A* formulations, respectively. This is because the current density *J* takes the first derivative of *T* and the second derivative of *A*. To avoid spurious solutions on the current density, the element order used for discretizing *A* should be one order higher than that in *T* [17,25]. More detailed discussions on the meshing and solving element order of *T-A* models can be found in [14,25].

## B. Results and Discussions

AC transport currents at 50 Hz are imposed in every tape of the coil with 11 A amplitude. The distributions of normalized current density, magnetic flux density, and hysteresis losses are shown in **Figure 4**. To focus on validating the proposed model, we present the results from the *H* reference and the *T-A* SMSH models only. It can be seen that the *T-A* SMSH model successfully reproduces the current penetration and the distributions of magnetic flux density as the reference model presents. Comparing the hysteresis losses, one can find that the results of the five analyzed tapes correctly indicate the changing trends of the losses in every pancake, and agree with those calculated by the reference model very well.

The piecewise cubic Hermite interpolating polynomial (PCHIP) method is used to obtain the total loss of the coil [14]. The total losses, degrees of freedom (DOFs), computing time, and efficiencies of all the numerical models are shown in **Table 3**.

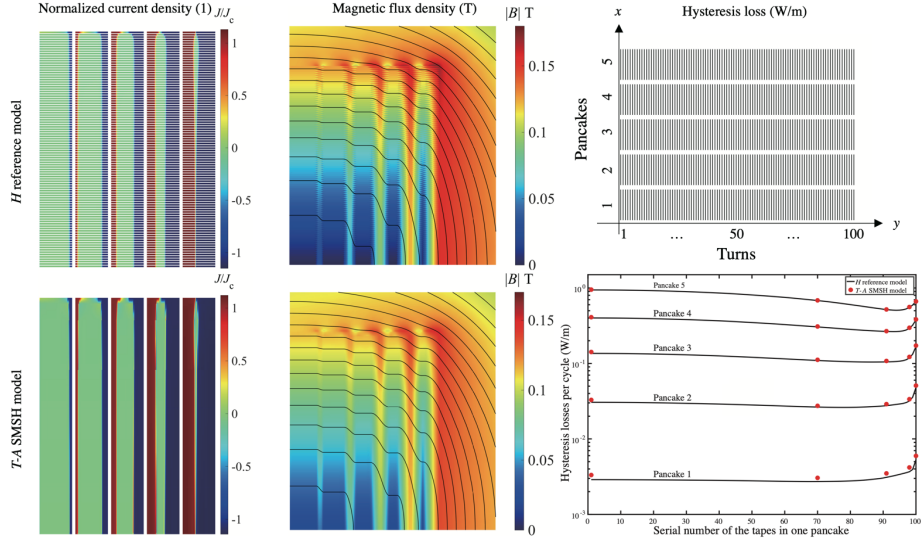

**Figure 4**. Distributions of normalized current density, magnetic flux density and hysteresis losses of the *H* reference model and the *T-A* SMSH model ($I_m$=11 A, at 15 ms).

**Table 3**. The total losses, DOFs, computing time, and efficiencies of different models in the case of 11 A.

| Models | Hysteresis loss[a] (W/m) | Degrees of freedom | Computing times[b] (h) | DOFs solved per second |
|---|---|---|---|---|
| *H* reference | 126.58 | 538918 | 25.59 | 5.8 |
| *H* full | 128.50 | 204111 | 8.96 | 6.3 |
| *T-A* full | 127.36 | 170231 | 0.50 | 94.6 |
| *H* homogeneous | 127.43 | 12551 | 0.18 | 19.4 |
| *T-A* homogeneous | 128.35 | 30685 | 0.087 | 98.0 |
| *T-A* SMSH | 125.98 | 17643 | 0.054 | 90.8 |

[a] The loss is the result of 1/4 part of the whole 2000-turn coil system.
[b] All the models are calculated on the same desktop with CPU i7-8700 @ 3.2 GHz, and RAM 32 GB.

From **Table 3**, we can find that the total hysteresis losses calculated by different models are highly consistent with the most significant error of 2.0%. And the error between the results from the proposed *T-A* SMSH model and the *H* reference model is 0.47%. Moreover, compared with the losses calculated by other authors: 127.02 W/m [18], 127.48 W/m [23], and 127.24 W/m [17], the high accuracy of the proposed model on calculating hysteresis loss is validated.

The computing efficiency of a model can be defined as the number of DOFs solved per second. One can see that the efficiencies of the three *T-A* models are at about the same magnitudes, and are much higher than those of the *H* models. Combining the homogenization technique, the *T-A* homogeneous and the *T-A* SMSH models are the two most time-saving models, both of which can finish the simulation within several minutes. Since the coil in the *T-A* SMSH model is divided by the analyzed tapes, with fewer bulks come fewer DOFs to be solved. Thus, the computing time of the *T-A* SMSH model is shorter than that of the *T-A* homogeneous model. It can be expected that the computing time will be further reduced, if the *T-A* homogeneous model is built with fewer bulks [17]. However, the accuracy should be carefully considered with reduced bulks since the bulk domains solve the HTS electromagnetic variables directly. For example, if the *T-A* formulation is applied in the bulk domains divided as the *T-A* SMSH model, the computing time reduce to 0.060 h. However, the result becomes inaccurate, which is presented by the instantaneous losses shown in **Figure 5**.

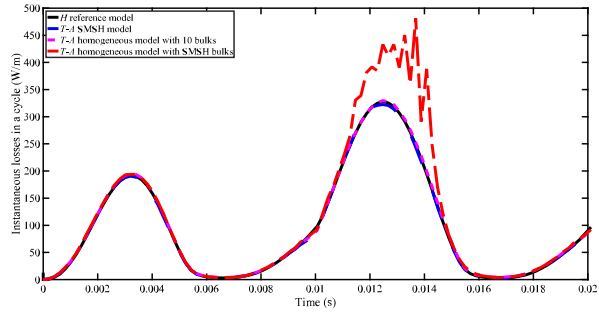

**Figure 5**. Instantaneous losses calculated by different models. The meshing elements along the HTS tapes' widths are 60 in all the *T-A* models.

In addition, the scenario of the coil carrying the current with 28 A amplitude is also simulated by the *T-A* simultaneous multi-scale homogeneous model. The distributions of normalized current density and magnetic flux density are shown in **Figure 6**, which are consistent with the results in [18] (referring to Figure 11). Hysteresis loss calculated by the proposed model is 925.48 W/m, whose error is 0.91% compared with 933.99 W/m from the reference model in [18].

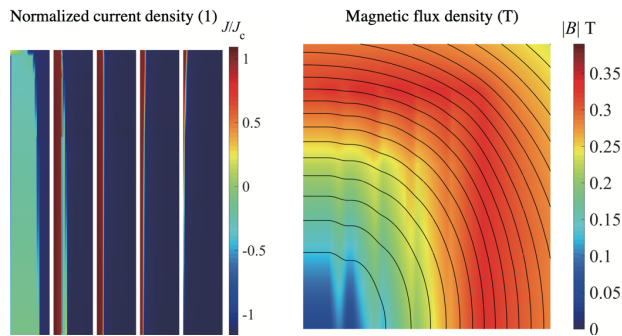

**Figure 6**. The distributions of normalized current density and magnetic flux density of the coil in the case of 28 A, at 15 ms.

## IV. APPLICATION IN 3D SIMULATIONS

Furthermore, we extend the *T-A* simultaneous multi-scale homogeneous method to simulate

3D HTS coil models. The detailed governing equations of the *T-A* formulation in 3D simulations can be found in [11].

**A. Modeling description and Case study**

A racetrack coil with 50 turns is simulated, which was studied by 3D *H* homogeneous method [9] and multi-dimension method [21]. The detailed geometric parameters are shown in **Figure 7**.

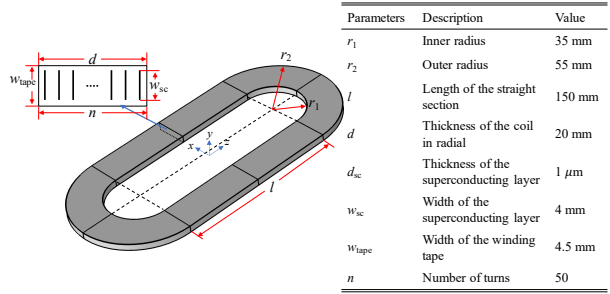

**Figure 7**. Sketch of the racetrack coil and geometric parameters.

Based on the symmetry, only 1/8 racetrack coil model is simulated with boundary conditions. Within the 50 tapes numbered from inside out, 10 tapes with the serial number of {1 2 4 8 16 35 43 47 49 50} are selected and built as HTS tapes with thin strip approximation, shown in **Figure 8**.

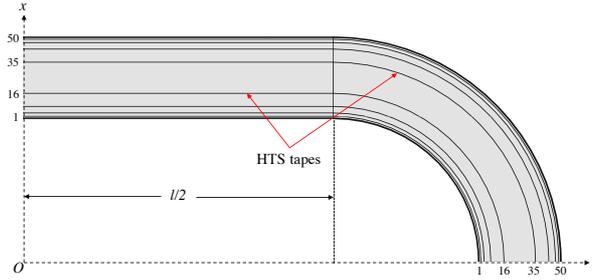

**Figure 8**. The racetrack coil geometry built in the 3D *T-A* SMSH model.

The 10 analyzed tapes are solved by *T* formulation to obtain the current density. And equations (4), (5) and (6) are used to estimate the current densities in the 3D bulk domains. The independent variables of the linear functions $F_1$ and $F_2$ are determined as $x$ and $\sqrt{x^2 + (z - l/2)^2}$, corresponding to the cases of $z \leq l/2$ and $z > l/2$, respectively. General Extrusion operators call the current densities in the analyzed tapes. Likewise, the magnetic field of the racetrack coil is calculated by *A* formulation with the current densities in the bulk domains only.

Equations (9) and (11) are used to describe the electromagnetic characteristics of the tapes in the racetrack coil. The detailed parameters and the expressions of magnetic flux density components can be found in our previous study [21]. 15 elements are meshed along the 2-mm half-width of the tapes. The *T* and *A* formulations are solved with linear and quadratic elements, respectively. The hysteresis losses in the analyzed tapes are calculated by

$$Q[\text{J/cycle}] = 2 \int_{T/2}^{T} \int_{S} \left[ \left( E_x \cdot J_x + E_y \cdot J_y + E_z \cdot J_z \right) \cdot \delta \right] dS dt \qquad (12)$$

where $S$ is the surface area of a thin-strip analyzed tape, and $\delta$ is the thickness of a tape. The losses in the non-analyzed tapes are estimated by interpolation.

**B. Results and Discussions**

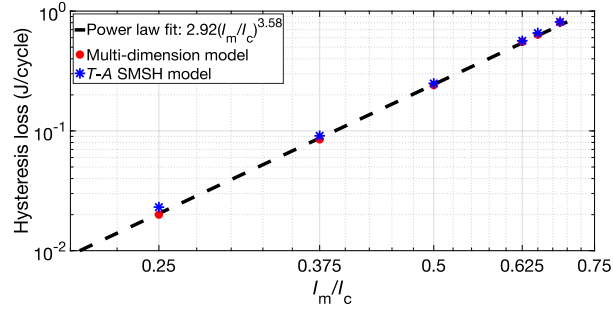

**Figure 9**. Hysteresis losses calculated by the 3D *T-A* SMSH model, compared with the results from the multi-dimension model and a power law for the 3D *H* homogenized model.

The racetrack coil is simulated in the cases of 50 Hz alternating transport currents with different amplitudes. The hysteresis losses calculated by the 3D *T-A* SMSH model are compared with the results obtained by the multi-dimension model [21] and the power law fit for the 3D *H* homogenized model [9], shown in **Figure 9**. $I_m$ is the amplitude of the transport current, and $I_c$ is the self-field critical current of the tape (160 A). One can find that the losses from the proposed and the previous models are in good agreement with the largest error of 13.2% in the case of $I_m = 0.25 I_c$. The error reduces to 9.3% if the meshing elements along the 2-mm half-width of the tape increases to 30 in the 3D *T-A* SMSH model, but the solved DOFs and the computing time increase significantly. The meshing elements should be appropriately selected to balance the computing accuracy and time, especially in the 3D model where quadratic-order elements are solved. Besides, we can find that the proposed model reliably predicts the hysteresis losses with 15 elements meshed along half-width of the tape. Therefore, the computing time is counted based on this meshing structure.

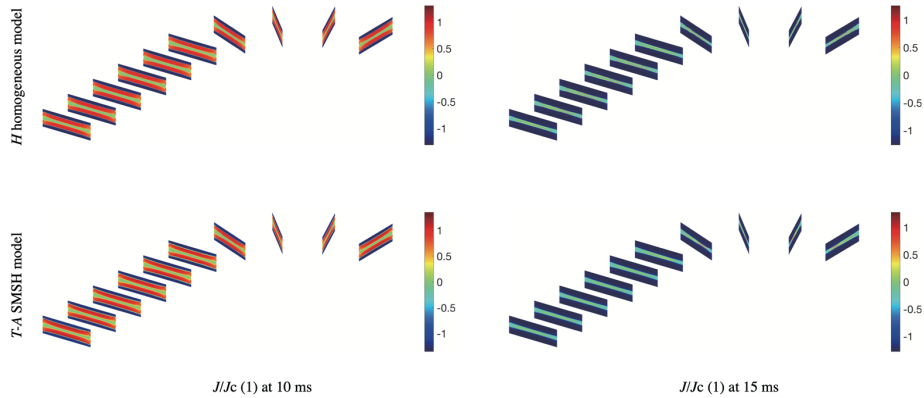

**Figure 10**. Normalized current density distributions of the racetrack coil at different times. Results from the 3D *H* homogenized model and the 3D *T-A* SMSH model are presented.

The normalized current densities obtained by the 3D *T-A* SMSH model at different times are presented and compared with those from the 3D *H* homogeneous model in the case of $I_m = 100$ A, shown in **Figure 10**. Specifically, **Figure 11** presents the normalized current density distributions at two particular planes: the middle of the straight section and the middle of the round section. It can be seen that the characterizations of the current penetration are presented clearly by the 3D *T-A* SMSH model, and coincide with those from the 3D *H* homogeneous model very well.

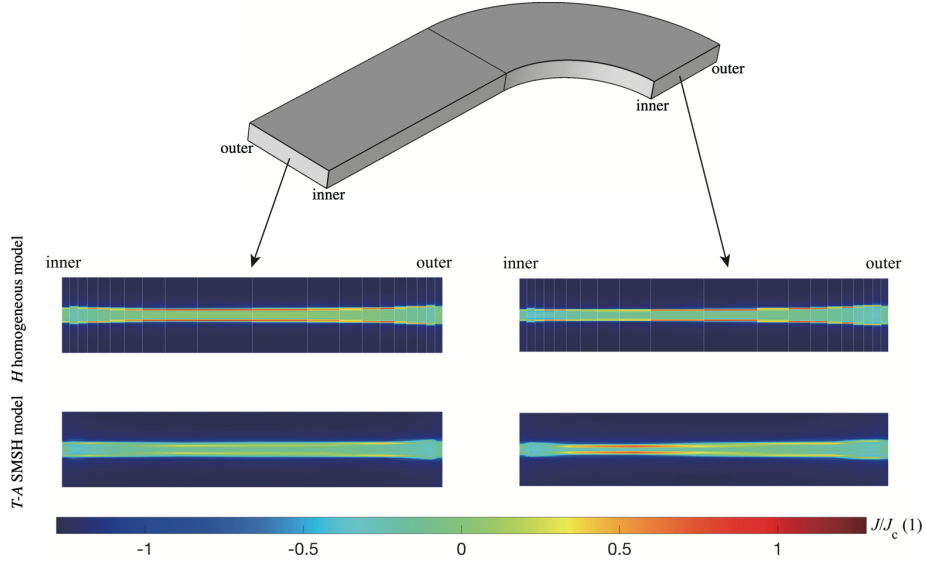

**Figure 11**. Normalized current density at 15 ms for the middle of the straight section (left) and the middle of the round section (right).

In the multi-dimension method study, we compared its computing time of simulating the racetrack coil with the 3D *H* homogeneous model in the case of $I_m$ = 100 A, where the 3D *H* homogeneous and multi-dimension models spent 8.6 h and 2.9 h, solving 238505 and 306911 (*coil submodel*) + 679 (*single-tape submodel*) DOFs, respectively [21]. Here, we compare the computing time of the 3D *T-A* SMSH model with the multi-dimension model directly, which is listed in **Table 4**.

**Table 4**. Computing time[a] of the 3D racetrack coil simulation by multi-dimension model and *T-A* SMSH model.

| $I_m$ (A) | Multi-dimension model | | *T-A* SMSH model |
|---|---|---|---|
| | Number of iterations | Computing time | (263196 DOFs) |
| $0.250I_c$ | 4 | 3.18 h | 0.88 h |
| $0.375I_c$ | 4 | 3.29 h | 1.34 h |
| $0.500I_c$ | 3 | 2.81 h | 1.89 h |
| $0.625I_c$ | 3 | 2.92 h | 2.57 h |
| $0.650I_c$ | 3 | 2.94 h | 2.69 h |
| $0.750I_c$ | 3 | 3.07 h | 3.32 h |

[a]All the models are calculated on the same desktop with CPU i7-8700 @ 3.2 GHz, and RAM 32 GB.

As the applied current gets close to the critical current, it takes more time to complete the simulation by the *T-A* SMSH model. The iterative method was employed for the multi-dimension model to enhance the calculation accuracy [21], while the number of iterations mainly determines the computing time to meet the criteria. Since the initial distribution of uniform current density is becoming similar to the current penetration in superconductors, fewer iterations are needed. Thus, the computing time at higher currents is less than those at lower. In contrast, the time is still longer at larger currents in the cases with the same iteration number. Therefore, one can find that the 3D *T-A* SMSH model is much more time-saving than the multi-dimension model for simulating the coil with small currents, and the computing time of both models is comparable with the currents increasing. It should be noticed that the computing time of the multi-dimension model is counted with the total time of all the analyzed tapes calculated in series, and the time can be reduced if parallel calculations are employed [18,21]. Although the proposed model with the simultaneous multi-scale method cannot use the parallel calculation, the solving process is more straightforward since only one computing file is needed. Moreover, the computing efficiency is encouraging too. Both models are recommended for simulating large-scale HTS coils.

## V. CONCLUSIONS

In the present paper, we propose an ingenious idea to combine the thin-strip, homogeneous and multi-scale methods in one numerical model. Innovatively, we treat the bulks' boundaries and domains with different electromagnetic properties, and solve them by $T$ and $A$ formulations, respectively. The study focuses on introducing the modeling process and benchmarks of the combination method. It has been demonstrated that the proposed model fully employs the advantages of the three advanced methods, which could rebuild the target coils with extremely simplified geometries. And it has been validated that the solving of the proposed model is highly accurate and time-saving both in the 2D and 3D simulations. However, due to the relationships of the current density and the current and magnetic potentials, the $A$ formulation should be solved with quadratic orders at least to maintain the current density distribution accurate, which results in large numbers of DOFs to be solved in the 3D $T$-$A$ models if the meshing is fine. The problem also occurs in the 3D $T$-$A$ SMSH model and should be studied carefully later.

We conclude that the proposed model provides a feasible approach to simulating coils with a huge number of stacked tapes. And we will continue exploring its applications in solving practical HTS systems with complex geometries.

**ACKNOWLEDGMENTS**

This work is supported by the SKP of China (Grant Nos. 2016YFA0300504 and 2017YFA0304204), the NSFC Grant No. 11625416, the Shanghai Municipal Government (Grants Nos. 19XD1400700 and 19JC1412702), the China Postdoctoral Science Foundation (Grant No. 2019M661355).

**AUTHORS DECLARATIONS**

**Conflict of Interest**

The authors have no conflicts to disclose.

**DATA AVAILABILITY**

The data that supports the findings of this study are available within the article.